%% file: main.tex
\documentclass[num-refs]{nbdt-article}

\usepackage{siunitx}
\usepackage{amsmath, amsfonts}
\usepackage{booktabs} %
\usepackage{hyperref}
\usepackage{amsmath}
\usepackage[T1]{fontenc}
\usepackage{lmodern}

\begin{document}

\title{
    Reading Decisions from Gaze Direction during Graphics Turing Test of Gait Animation
}

\author[1]{Benjamin Knopp}
\author[1]{Daniel Auras}
\author[1]{Alexander C. Schütz}
\author[1]{Dominik Endres}
\affil[1]{%
    Department of Psychology\\
    University of Marburg\\
    Germany\\
    Marburg
}

\corraddress{Benjamin Knopp}
\corremail{benjamin.knopp@uni-marburg.de}
\runningauthor{Knopp et al.}

\maketitle

\begin{abstract}
    We investigated gaze direction during movement observation.
    The eye movement data were collected during an experiment,
    in which different models of
    movement production (based on movement primitives, MPs) were compared in a
    two alternatives forced choice task (2AFC).

    Participants observed side-by-side presentation of two
    naturalistic 3D-rendered human movement videos, where one video was based
    on motion captured gait sequence, the other one was generated by
    recombining the machine-learned MPs to approximate the same movement.
    The task was to discriminate between these movements while their eye movements were recorded.
    We are complementing previous binary decision data analyses with eye tracking data.
    Here, we are investigating the role of gaze direction during task execution.
    We computed the shared information between gaze features and decisions of the participants, and between gaze features and correct answers.

    We found that eye movements reflect the decision of participants during the 2AFC task,
    but not the correct answer.
    This result is important for future experiments, which should take advantage of eye tracking to complement binary
    decision data. 

\end{abstract}

\section{Introduction}

Knowledge about the human perceptual system relies to a large degree on
psychophysical experiments, where participants respond to visual stimulation
with a binary answer, e.g. button presses to investigate perception of terrain
\cite{scottEvaluatingRealismExampleBased2022}, High Dynamic Range videos
\cite{hindeImmersivePropertiesHigh2022} and triangle meshes
\cite{almutairiImperceptibilityThresholdsQuantised2021}.
One particularly famous and attractive paradigm is the two-alternative
forced choice task, dating back to Fechner \cite{fechnerElementePsychophysik1860},
where participants are forced to choose between two alternatives. The
appeal is, that response biases can be reduced if one alternative serves as the
ground truth baseline.

While the awareness of the importance of rich, naturalistic stimulus displays steadily increases
\cite{fieldRelationsStatisticsNatural1987,rideauxStillItMoves2020} , the
corresponding recorded behavioural response is typically still impoverished.

One example of such an experiment is a previous study
\cite{knoppPredictingPerceivedNaturalness2018}, which utilize
naturalistic videos of human gait. The goal of the study was to find out
if movement based on primitives (more details in
\cite{knoppPredictingPerceivedNaturalness2018}) are able to meet the
perceptual expectations of the observers. The experiment used a variant of the graphics
Turing test \cite{mcguiganGraphicsTuringTest2006}, in which participants where
presented a "real" movement on one side of the display, and an "artificial" one
on the other side. Their task was to indicate by button press \emph{which movement they
perceived as more natural}.

The analysed behaviour thus corresponds to a binary, probably highly conscious
decision. Yet, the behavioural response to the stimulus is highly complex, involving
finger- and eye-movements that might tell us more about the decision process leading to the binary button press response.
Eye movements are strongly linked to perception, and do
tell us about the allocation of attention, which might inform about
the processing of stimuli and decisions made in order to complete the
experimental task. %

The authors of \cite{knoppPredictingPerceivedNaturalness2018} also recorded eye movements and
provided this data. The focus of this study is to re-evaluate the data, with a
focus on eye movements.

\section{Related work}

Gaze in the context of two alternatives forced choice tasks using
static stimuli has been investigated by several studies in the past
\cite{shimojoGazeBiasBoth2003, krajbichVisualFixationsComputation2010}.
\citet{shimojoGazeBiasBoth2003} let participants chose the more
attractive one of two faces. They observed the gaze cascade effect:
Fixations are biased towards the chosen stimulus, and this bias
is stronger for difficult tasks (faces with similar attractiveness
rating in their study) compared to easy tasks.
\citet{shimojoGazeBiasBoth2003} suggest that this effect implies active 
contribution of fixation to
preference formation, i.e. fixation bias is not just a consequence
of the decision.

A computational model of value-based decision implementing this assumption
has been proposed by \citet{krajbichVisualFixationsComputation2010}
and empirically tested in a free viewing 2AFC paradigm of participants
choosing images depicting a more desired food.
They model the decision process by a drift diffusion model, where (stimulus
value independent) fixations increase the drift towards the fixated
stimulus. This explains the last fixation bias, except in cases, where
the fixated stimulus depicts much averted food.

Similar to the experiment that our current study is based on, \citet{taubertOnlineSimulationEmotional2012,chiovettoPerceptualIntegrationKinematic2018}
use dynamic stimuli, but they did not analyse gaze behaviour.

Fixation patterns of static human posture were
studied by \citet{calbiVisualExplorationEmotional2021},
in a judgement of emotions paradigm.
Dynamic human movement stimuli were also investigated
in the context of emotion
in children
\cite{geanguLookBodyEyetracking2020}
and adults
\cite{moritaInfantAdultPerceptions2012a}.
\citet{mataricFixationBehaviorObservation1998a} investigated eye movement
during observation and imitation of body movement.

There are also studies with direct application in human-robot
interaction:
\citet{petterssonHumanMovementDirection2020,petterssonHumanMovementDirection2021}
used eye movements to predict movement direction, which is an
important computer vision problem to be solved in order to
make collaborative robotics a safe endeavour for humans.

\section{Methods}\label{methods}

In this study we focus on analysing gaze data recorded during an experiment, whose
decision data was already analyzed in
\cite{knoppPredictingPerceivedNaturalness2018}. Here we describe the additional
methods for collecting and analysing the eye movement data.

\hypertarget{experiment}{%
\subsection{Experiment}\label{experiment}}

\begin{figure}
\includegraphics[width=0.9\linewidth]{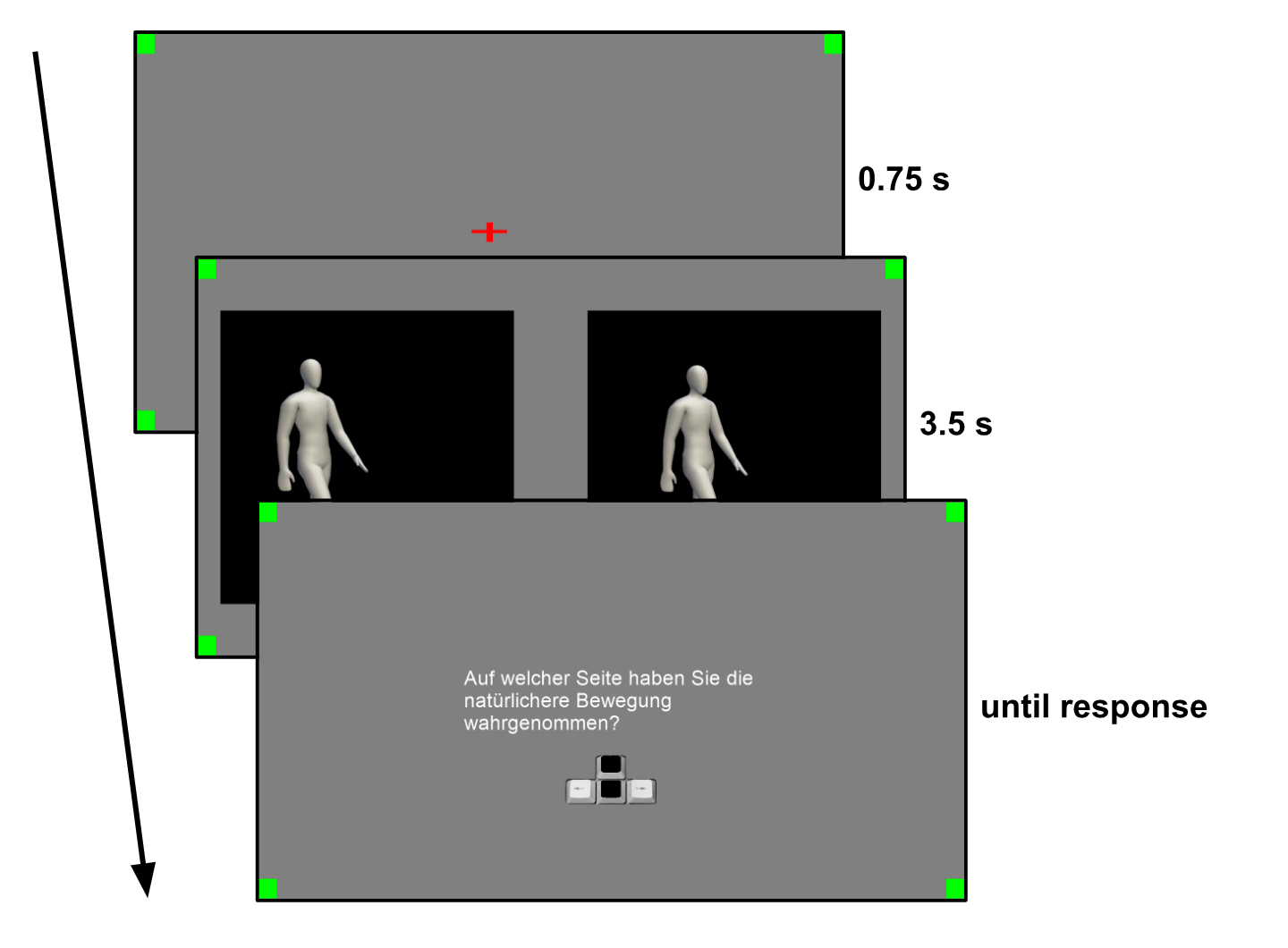}
\caption{
    Figure reprint with permission of \cite{knoppPredictingPerceivedNaturalness2018}:
    Illustration of experimental procedure. Each trial begun with a fixation period
    of 0.75s. Then, participants watched
    simultaneous replays of natural and generated movements for 3.5s.
    After the presentation the participants were asked
    "On which side did you perceive the more natural movement?"
    and responded using the arrow keys of a standard computer keyboard.
}
\label{fig:experimentgui}
\end{figure}

The experiment was designed as a graphics Turing test \cite{mcguiganGraphicsTuringTest2006}:
in each trial, movement primitive (MP) driven animations were displayed side-by-side
with a natural (mocap-based) animation. The participants were asked
to indicate which animation was perceived as more natural. The MP type and complexity
was manipulated for the artificial stimuli, to check how complex the model must
be to produce movement indistinguishable from real movement. For more details,
please refer to \cite{knoppPredictingPerceivedNaturalness2018}.

We have been kindly provided with additional data, i.e. nine participants,
who were wearing a mobile eye tracker to collect eye movement data. Each
participant completed 643 trials.

Data was recorded using
wearable SMI Eye Tracking Glasses 2w (SMI ETG) 60/120 Hz, which also
record the field of view with an integrated scene view camera 
(resolution: 1280x960p @ 24FPS,
field of view: 60° horizontal, 46° vertical). 
The device has a tracking accuracy of
0.5°.

The eye gaze in pixel coordinates of the scene view camera was exported by the
included SMI software.
We recorded eye movement with 60Hz and used time stamps to
select frames closest to the corresponding field of view frames
recorded with 24Hz.
The mobile and wearable device did not impose restrictions on head
movement of the partcipant.

\hypertarget{bodytracking}{
    \subsection{Tracking the keypoints of displayed avatars}\label{bodytracking}}

\input{DeepLabCut}

Neural network training, inference and data
preprocessing required a dedicated machine, running Ubuntu 18.04 LTS on an AMD
Ryzen Threadripper 2990WX 32-core processor @ 1716.724 MHz with 32 GB DDR RAM
memory, and a GeForce RTX 2070 8 GB GDDR6 graphics cards for hardware
acceleration.

\hypertarget{eyefeatures}{%
\subsection{Post calibration and gaze features}\label{eyefeatures}}

Wearable eye trackers provide much convenience for the experimenter
and comfort for the participant, yet this comes at the cost of
less robust and accurate calibration compared to static eye trackers.
We post-calibrated the eye tracker by the block-wise offset of the
gaze during inter stimulus interval towards the fixation cross
where necessary.

For saccade detection, we computed the speed of gaze direction, and set a threshold of
$100^\circ/s$. 
This is a conservative estimation based on \cite{raynerEyeMovementsReading}:
The authors report a duration of 50ms  for $5^\circ$ saccades. In our
experiment the avatars are approximately $16^\circ$ apart, resulting in
160ms (even though saccade speed increases for longer distances).

\begin{figure}[!ht]
    \centering
    \includegraphics[width=0.5\linewidth]{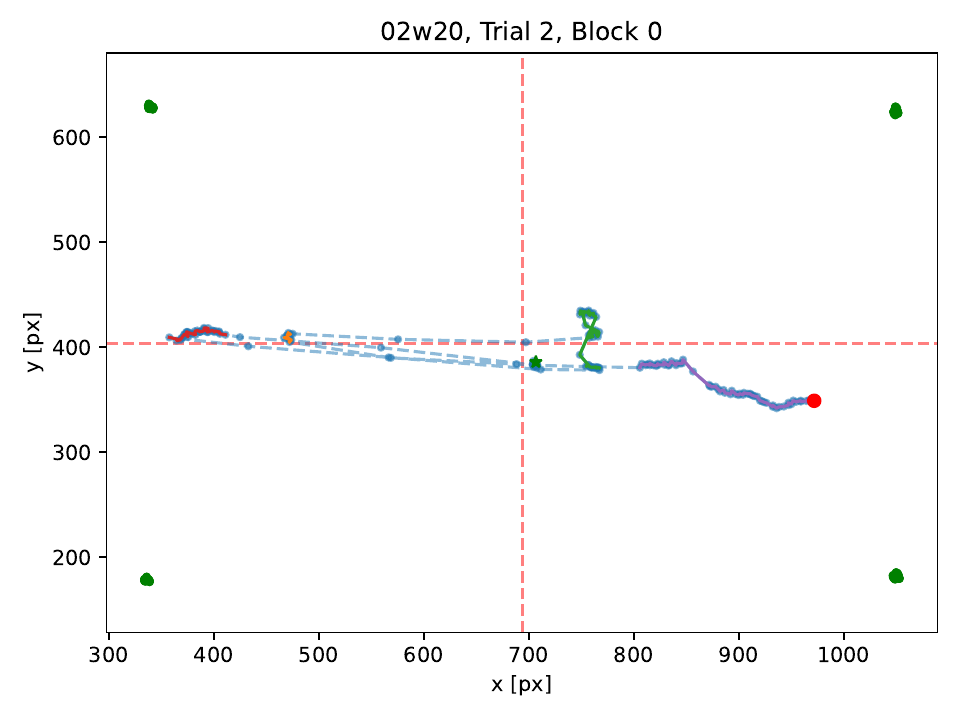}

    \includegraphics[width=0.5\linewidth]{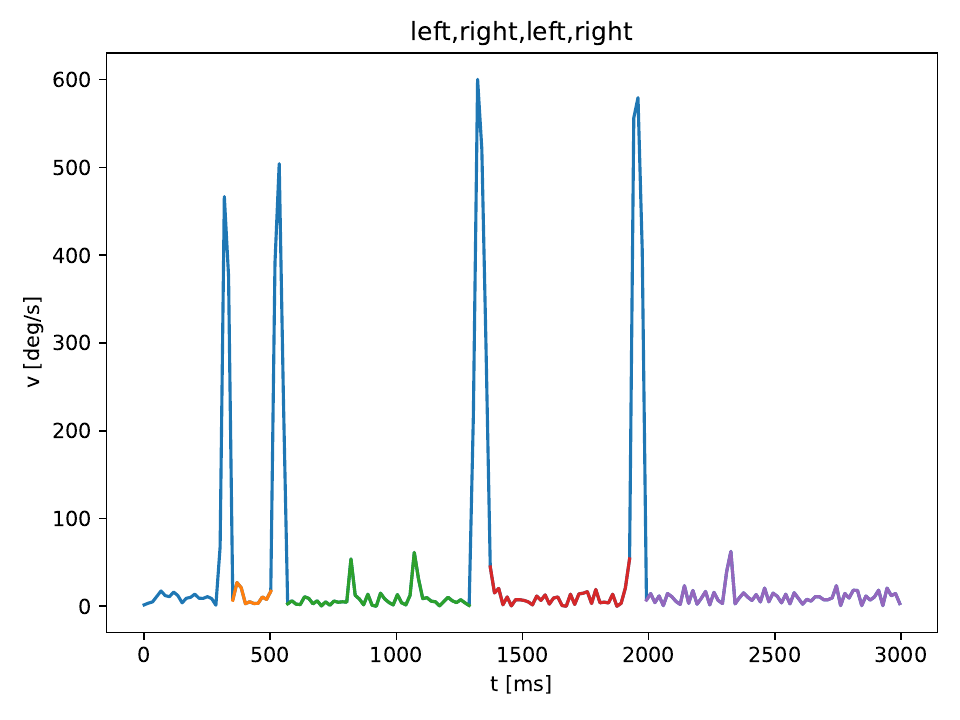}
    \caption{
        Upper pane:
        Gaze trajectory of one exemplary trial. The four green dots in the corners show the monitor
        corner marks (cf. fig. \ref{fig:labels}), and the fixation cross where the dashed red lines meet.
        The complete gaze trajectory is dashed blue, with small blue points marking the gaze
        direction in each frame. Each segment, where participants gaze follows the avatar
        is plotted as orange, green, red and purple lines.
        Below: Speed (angles / second) for the trajectory in blue. Speeds above $100^\circ/s$
        segment the trajectory, leaving the segments (color coded as above) of single avatar gazing.
        See text for more detailed description.
    }
    \label{fig:eyesample}
\end{figure}

This threshold yields sensible segmentation results, see Fig. \ref{fig:eyesample}:
In this trial, the participant makes a first saccade to the left side of the monitor (orange)
and quickly switches towards the right avatar, looking up and down its body (green).
Then she makes a saccade towards the left avatar again and smoothly follows
its central body trajectory (red). The last gaze is directed at the
right avatar again (violet)\footnote{In this trial, the left stimulus was
chosen, which was the artificially generated movement.}.
The
saccades crossing the sides to the other stimulus have speeds of above $100^\circ/s$.
These saccades are used to segment the gaze trajectory between the left and the right stimulus.
Please refer to the video in the supplementary material, which shows the scene video
recording with gaze direction corresponding to this trial, to get a better impression.

After checking the quality of the gaze data,
we manually extract several gaze features\footnote{We use the term gaze feature
because we use them in the context of logistic regression.} from it, based on
the eye tracker data and the keypoint label data:

\begin{itemize}
    \item \textbf{duration left/right}: duration of gazing at the left/right avatar
    \item \textbf{saccades}: saccades between left and right stimulus
    \item \textbf{first/last}: which avatar was gazed at first/last
    \item \textbf{upper-lower}: ratio of gazing at upper- vs lower body of the avatars
\end{itemize}

\hypertarget{logistic-regression}{%
\subsection{Logistic Regression}
\label{logistic-regression}}

Many variables of interest have a binary encoding: reported decisions, 
side of natural stimulus (task) and correctness of choice (uncertainty).
Logistic regression enables prediction of these variables (outcomes) for each
trial depending on continuous or discrete features.

We assume that the binary outcome of each trial $x_i$ is Bernoulli distributed
to have success with probability $\Theta$,
and investigate the relationship to $J$ features $y_{ij}$ with $j=1\dots J$.

\begin{figure}[!ht]
    \centering
    \includegraphics[width=0.5\linewidth]{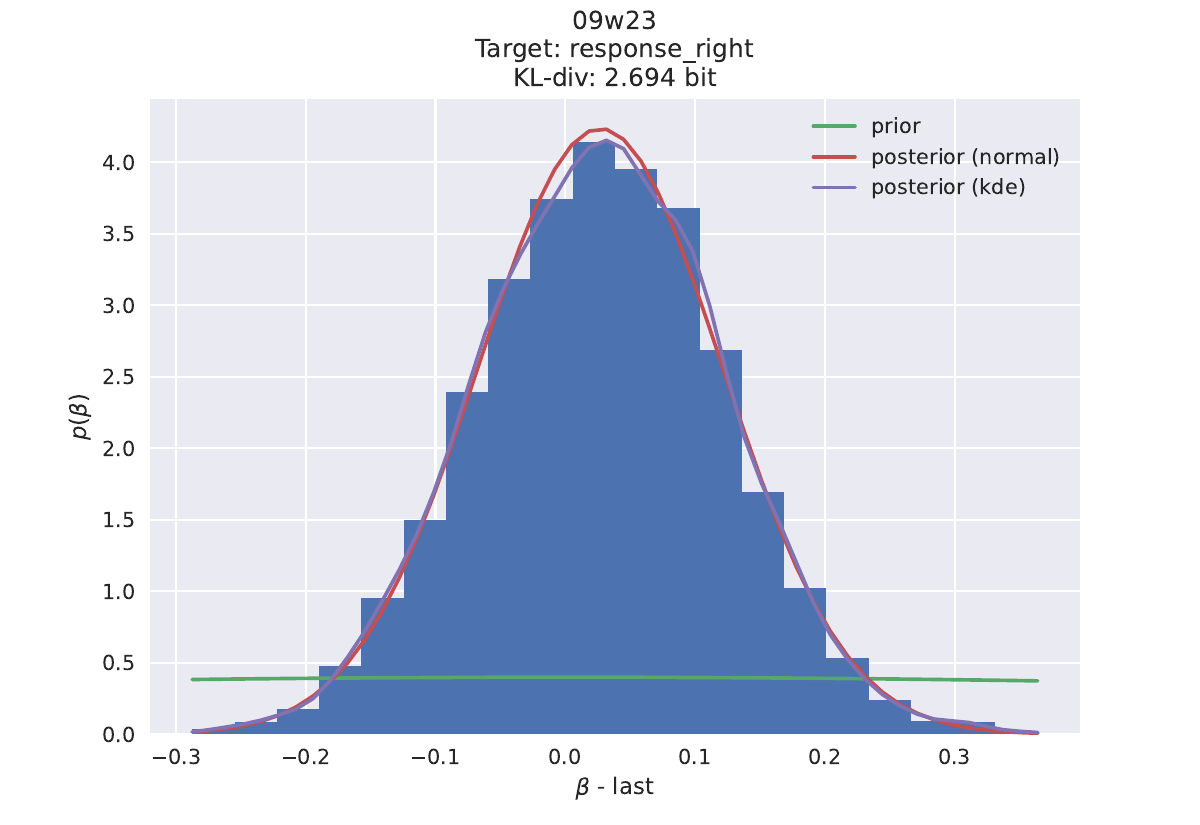}
    \caption{
        Histogram of posterior samples for model parameter $\beta_{\mathrm{last fixation}}$ to predict
        outcome variable response (decision). Violet line is the kernel-density estimation of the
        posterior distribution, red line is the normal distribution with parameters estimated from the samples.
        The normal is thus a very good approximation to the posterior. The green
        line shows the prior: $\mathcal{N}(0, 1)$, which seems almost uniform in comparison to the posterior.
        Thus, the data overrules the prior.
    }
    \label{fig:kldiv}
\end{figure}

We approximate the full posterior distribution of the model parameters $\alpha, \beta_j$ 
using Markov chain Monte Carlo\footnote{We use the No-U-Turn Sampler implemented in Python library
    PyMC3 \cite{salvatierProbabilisticProgrammingPython2016}.} given standard Gaussian priors. Please see
    Fig. \ref{fig:kldiv} which shows that the prior shape is overwritten
    by the data and check our code which will be available on publication).

\begin{align}
    x_i &\sim \text{Bernoulli}(\Theta_i)\\
    \Theta_i &= \frac{1}{1+\exp(-(\alpha + \sum_j\beta_j \cdot \text{y}_{ij}))}\\
    \alpha, \beta_j &\sim \mathcal{N}(0, 1)
\end{align}

The features $y_{ij}$ contain a lot of information about $x_i$ if the value of
$\Theta_i$ close to zero or one after including these features. Mutual
information, described next, measures how much information the $y_{ij}$ are
expected to deliver about $x_i$.

\hypertarget{explainmutinf}{%
\subsection{Mutual information}\label{explainmutinf}}

The mutual information $I(x, y)$  (see
\cite{coverElementsInformationTheory1999}) measures how much information is
shared between two
variables $x, y$ (for example $x$: trial outcome
and $y$: gaze feature). It is defined as the relative entropy
(a.k.a. Kullback Leibler divergence) between the joint distribution $p(x, y)$
of the variables and the product of their marginal distributions $p(x)p(y)$.

\begin{align}\label{eq:mutinf}
    I(x, y) &= \mathcal{KL}\left(p(x, y)|| p(x)p(y)\right)\\
            &= \int p(x, y) \log_2\frac{p(x, y)}{p(x)p(y)} dx dy
\end{align}

Intuitively, this tells the difference it makes when we do not assume independence
(no information shared): From $p(x, y) = p(x)p(y)$ follows $I(x, y)=0$.

Simple reformulation of Eq. \ref{eq:mutinf} shows that mutual information can also be
written as the difference of entropy $H(x)$ and the conditional entropy $H(x|y)$:

\begin{align}\label{eq:mutinf2}
    I(x, y) &= H(x) - H(x|y)\\
            &= \int p(x)\log_2 p(x)dx - \int p(x|y)p(y)\log_2 p(x|y)dxdy
\end{align}

This shows us that the mutual information is maximal if $y$ removes
the uncertainty about the value of $x$ completeley, i.e. $p(x|y)$ is zero or one in case
of discrete variables, then the conditional entropy is zero. In this study, we
investigate mutual information where at least one variable is binary.
Thus, we expect a maximum $I(x,y)$ of one bit.

While $p(x)$ and $p(y)$ are given from the experiment, $p(x|y)$ is estimated using
logistic regression (see Section \ref{logistic-regression}).

\hypertarget{modelevidence}{%
\subsection{Model evidence}\label{ch:modelevidence}}

The model evidence, a.k.a. marginal likelihood, is the central quantity for 
Bayesian model comparison: It describes the likelihood of the data $\mathcal D$
marginalized over model parameters $w$ for a specific model $M$ (for example
specified by the set of features). The model evidence is small if the model
is not able to replicate the data (due to $p(\mathcal D|w, M)$), but also if 
the model is too complex (because $p(w|M)$ is too spread out), therefore
implementing Occam's Razor.

\begin{align}
    p(\mathcal D|M) &= \int p(\mathcal D|w, M)p(w|M) dw\\
         &= E_{p(w|M)}[p(\mathcal D|w, M)]\label{eq:modelevidence}
\end{align}

Typically, the integral has no analytical solution and must be approximated.
In this study we use bridge sampling \cite{gronauTutorialBridgeSampling2017},
which tends to yield robust approximations from posterior samples.

\hypertarget{results}{%
\section{Results}\label{results}}

We present our results below, making reference to the methods section
(Section \ref{methods}) where appropriate.

\hypertarget{eyemovementstatistics}{%
\subsection{Eye movement statistics}\label{eyemovementstatistics}}

\begin{figure}[!ht]
    \centering
    \includegraphics[width=1\linewidth]{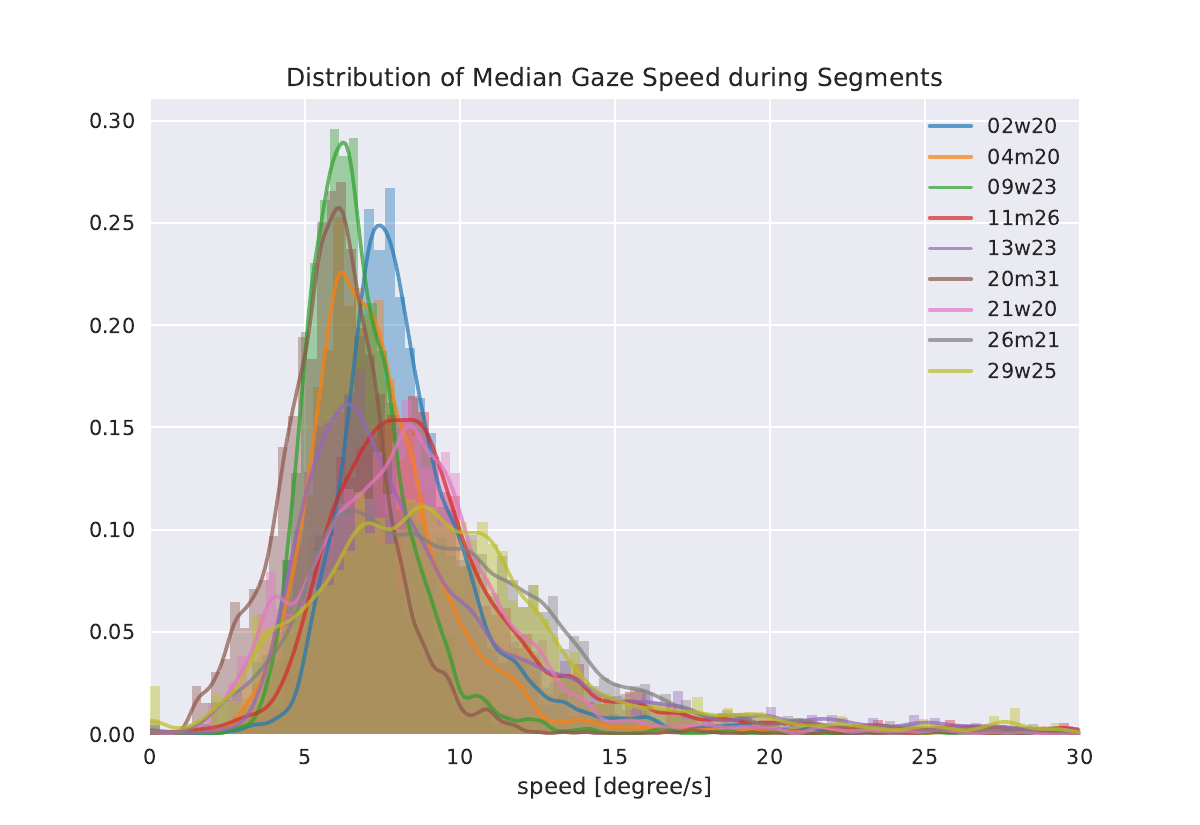}
    \caption{
        Gaze speed distribution. Bars show normalized bin counts, lines a Gaussian kernel
        density estimation in corresponding color. Color codes participant ID.
    }
    \label{fig:gazespeed}
\end{figure}

We computed the gaze speed distribution, see Fig. \ref{fig:gazespeed}.
Most participants show a unimodal distribution peaking at 6 to 8 degree
per second. This is higher than the average speed of the avatar of
about 3 degree / s, which indicates that these participants do not
smoothly pursue the avatar during a segment, but rather make small
saccades within the avatar. 

\hypertarget{mutinf}{
    \subsection{Gaze features share information with decision}
    \label{fig:mutinf}
}

We computed the mutual information (see Section \ref{explainmutinf}) between
several gaze features $y$ (gaze duration left and right, number of 
between-stimulus-saccades,
first and last fixation, see Section \ref{eyefeatures}) and three different
variables $x$: The decision, the task (i.e. location of the natural animation),
and if the choice was correct.
The result is shown in Fig. \ref{fig:mutinf}.

\begin{figure}[!ht]
    \centering
    \includegraphics[width=1\linewidth]{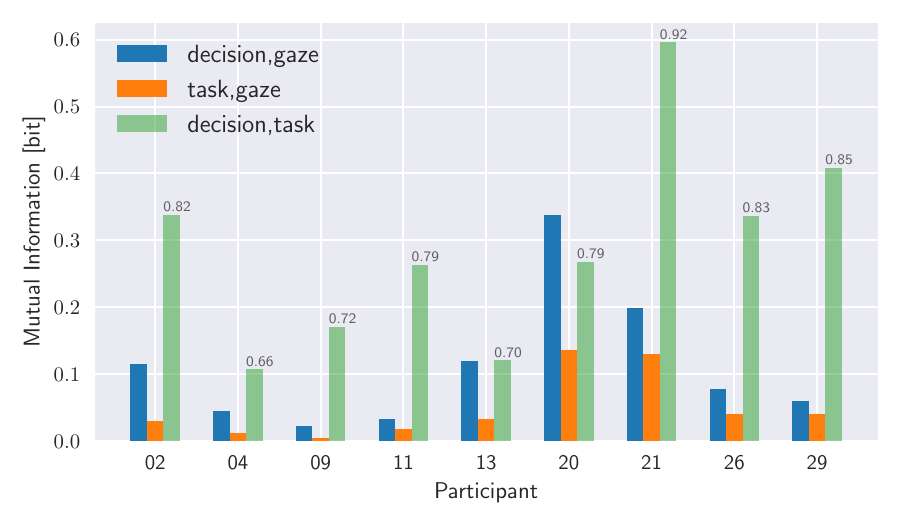}
    \caption{
        Mutual information between decision and gaze (blue), task and gaze (orange)
        and decision and task for all participants. The scale is the same for all bars, the number above green bar is the mean of correct answers to illustrate the connection between mutual information between task and decision and participant performance.
    }
    \label{fig:mutinf}
\end{figure}

For all participants, the gaze features share more information with the button pressed in
comparison to the location of the natural stimulus. Gaze features do
not share information about the choice being correct (indistinguishable from
zero, omitted from plot). As expected, the mutual information between decision and task
reflects the performance of the participants.

\hypertarget{featureimportance}{%
\subsection{Relative importance of gaze features for predicting decisions}\label{featureimportance}}

\begin{figure}[!ht]
    \centering
    \includegraphics[width=1\linewidth]{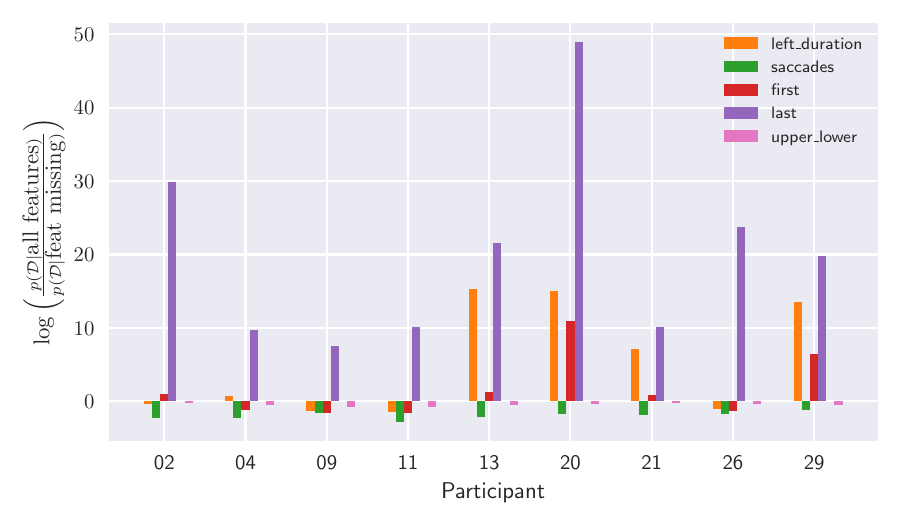}
    \caption{
        Feature Importance: Log-odds of between marginal likelihood of
        model with all features and model with one feature missing.
        The missing feature is one of the six bars shown for each
        participant. Positive values indicate that leaving out the
        particular feature decreases the likelihood, i.e. the
        feature is important for decision prediction.
    }
    \label{fig:feature-importance}
\end{figure}

After establishing the closer connection of gaze to decisions compared to natural stimulus location,
we further investigated the logistic regression model for decisions.
To estimate the relative predictive importance of each gaze feature, we carried out a leave one feature out analysis:
The model was trained multiple times, with one feature left out during each training run. To compare
the results, we estimated the model evidence (see \ref{ch:modelevidence}), and subtracted the model evidence of the model
with full feature set. This yields a log odds ratio
of how much more likely the full model is relative to the model with the investigated feature missing.
The result is shown in Fig. \ref{fig:feature-importance}.

For all participants, the last fixation the most important feature.
Next to that, the duration of looking at one side is important for four
participants (13, 20, 21, 29).
The first fixation increases model evidence for two participants (20 and 29).
All other features are not relevant, the number of saccades and the gaze
ratio between upper- and lower body even slightly decrease model evidence
for all participants (introducing unnecessary complexity).

\hypertarget{cascade}{%
\subsection{Gaze cascade effect}\label{cascade}}

\citet{shimojoGazeBiasBoth2003} introduce the gaze cascade effect:
more difficult tasks increase the last fixation bias (i.e. the tendency
of choosing the stimulus which was fixated last).
We wanted to determine if this bias is present in our data too. If that was the case, then the task difficulty
should be predictive of the congruence between last fixation and reported decision.
We therefore inferred a logistic regression model of that congruence using the crossvalidatory
mean squared error (MSE) between natural movement and MP generated movement as a predictor
(see \cite{knoppPredictingPerceivedNaturalness2018} for details).
There, it was shown that MSE provides a good proxy for perceived
naturalness: The lower the MSE, the higher the perceived naturalness of
the model generating the artificial movement. It thus provides an
approximation for task difficulty.  We operationalise this as $\beta_{MSE}<0$,
and report posterior  $p(\beta_{MSE}<0)$ in the table below.

\begin{table}[!ht]
\centering
    \caption{Probability of gaze cascade effect given participant data.}
\input{tables/cascade.tex}
    \label{tbl:cascade}
\end{table}

\newpage
Our results here are inconclusive, there are three participants
showing the gaze cascade effect (02w20, 09w23, 26m21), two which do not show it (04m20, 29w25), and four
where the result is uncertain (11m26, 13w23, 20m31, 21w20) (See Table \ref{tbl:cascade}).

\hypertarget{discussion}{%
\section{Discussion}\label{discussion}}

In this study, we investigated the fixation patterns during a
naturalness discrimination two alternatives forced choice task of human whole
body movements.

We found, that gaze features do tell us more about the decision
of the participant than about the actually correct answer (see Fig. \ref{fig:mutinf}). 
This result might be due to the task instruction to "choose the more 
natural stimulus", which leaves plenty of room for interpretation.

The gaze features might also inform us about the uncertainty
of the participants. In this case we would expect, that
gaze features and correctness of answer would share information.
We have not found this to be the case. This is surprising,
because we included the number of gaze shifts between avatars
as feature. Increasing the number of shifts might be a plausible
strategy to gain more information in case of difficult decisions.
One reason why this strategy might not have worked is the fixed
length of 3.5s stimulus display.

Given the results of the mutual information analysis, we
investigated the importance of gaze features we used
(see Fig. \ref{fig:feature-importance}) to predict participants' decisions.
The last fixation is for all participants by far the most important
feature: Participants co-allocate their last fixation with their
decision. This is in line with the last fixation bias observed
in 2AFC paradigm with static stimuli 
\cite{krajbichVisualFixationsComputation2010, shimojoGazeBiasBoth2003}.

The first fixation does not provide a large contribution for predicting the
decisions, except for two participants (Fig. \ref{fig:feature-importance}): the first fixation might be
driven to a large degree
by the direction of the walking movement, which was randomly
mirrored. If the walking direction was from left to right,
the distance from fixation cross to the right avatar was far
shorter than to the left (and vice versa) \cite{araujoEyeMovementsVisual2001}.
"Lazy" first fixations
would therefore be not informative for decisions due to randomization.

We have checked for the existence of gaze cascade effect, and found
inconclusive results: congruence between last fixation and decision was predictable from
task difficulty for only two participants.

The experiment was designed mainly with focus on comparing different
movement production models for their viability as perceptual
representation. Therefore, our study has some limitations:
We computed the distribution of gaze speed during segments (Fig.
\ref{fig:gazespeed}). \citet{agtzidisBetterUnderstandingEye} found pursuit
speeds ranging from three to nine degree per second (between first and third
quartile) in free viewing naturalistic video clips, which is comparable
to our study. Yet, the speed of the avatar was fixed. In future experiments
focusing on analysis of eye movements during naturalistic movement,
more diverse angular speeds of movements should be used. This
could be achieved by continuous change of perspective compared to
simple mirroring of the movements. This additional randomization
is important, because of the suggested causal role of fixations on
decisions \cite{shimojoGazeBiasBoth2003, krajbichVisualFixationsComputation2010}.

\hypertarget{conclusions}{%
\section{Conclusions}\label{conclusions}}

Our study provides insight about the role of eye movements
during human movement observation 2AFC task: we provide evidence
that gaze features are closely related to the decisions
of the participants. We propose that the relation of eye
movements to the task of naturalness discrimination is
mediated by the reported decision of the participant.

\section*{Code and Data Availability}
The code and data is available at \href{https://gitlab.uni-marburg.de/knoppbe/movementprimitivegaze}{gitlab.uni-marburg.de/knoppbe/movementprimitivegaze} and
\href{https://tam-datahub.online.uni-marburg.de/entities/dataset/ca3a1cd7-a268-4ec9-8898-30885c36c903}{tam-datahub.online.uni-marburg.de}.

\section*{Acknowledgements}
\thispagestyle{empty}
 This work was funded by DFG, IRTG1901 - The brain in action, and SFB-TRR
 135 - Cardinal mechanisms of perception. We thank Olaf Haag for help
 with rendering of the stimuli and collecting data, and Dmytro Velychko for MP models.

\bibliography{DeepLabCut-references.bib, references.bib}
\end{document}

%% file: DeepLabCut.tex
To make sense of the gaze data provided by the wearable eye tracker, it was
neccesary to relate the raw pixel
coordinates to the animated movement. 
We label keypoints of the avatars' body: feet, knees, pelvis, hands, elbows,
shoulders, torso, manubrium and head. These keypoint labels enable
robust perception of movement in humans \cite{johanssonVisualPerceptionBiological1973} and 
should therefore enable calculation of meaningful gaze features relative to these markers (see next section).
In order to track
body coordinates, we deployed a pre-trained deep neural network (DNN) model provided by the
Python library \texttt{DeepLabCut} \cite{Mathisetal2018, NathMathisetal2019},
which was originally developed for determining poses of laboratory animals.

\begin{figure}[!ht]
\centering
\includegraphics[width=0.5\linewidth]{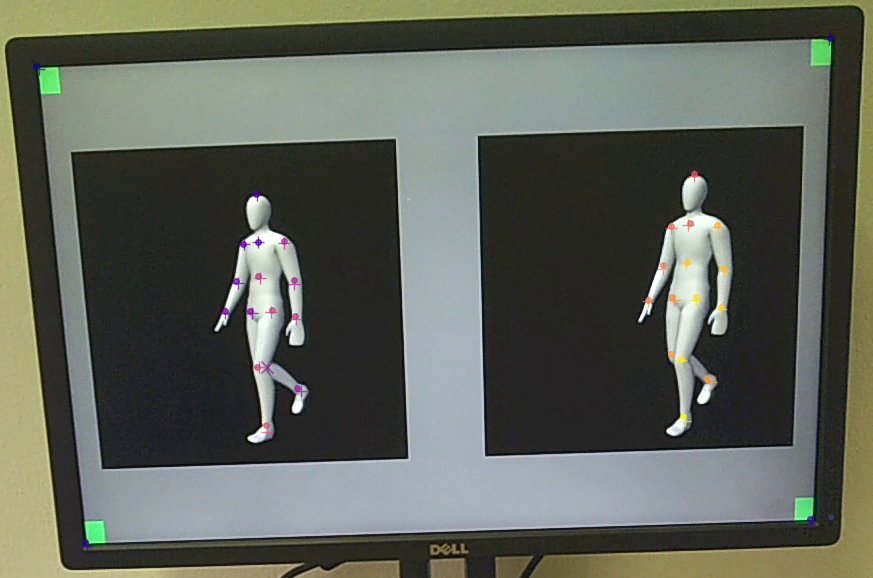}
\caption{
    Training labels are marked with "+", predictions by the model with
    "\(\cdot\)". Four green squares marked the display corners. Predictions
    with likelihood \(\leq p_{cutoff}\), are marked with "\(\times\)" (see
    occluded knee of left avatar).
}
\label{fig:labels}
\end{figure}

From the offered neural networks, we opted for a
ResNet-101 architecture \cite{he2015deep, insafutdinov2016deepercut}, because
it came pre-trained on human pose using the MPII Human Pose dataset
\cite{andriluka14cvpr} by way of \emph{supervised learning}. 
For fine-tuning the model to our dataset, we extracted a
total of 165 frames from the participants' point-of-view eye tracker recordings
(Fig. \ref{fig:labels}), based on the recommendation in \cite{Mathisetal2018}.

The frames were hand-picked, to incorporate intra-trial stimulus footage and to
be approximately uniformly distributed across the four blocks per experiment
run. Thereafter 34 labels were defined: four (green) squares mark the display corners and 15
per animated avatar, tracing their body movement (Fig \ref{fig:labels}). Every
frame was hand-labeled, whenever a label applied.  After fine-tuning on the set
of labeled frames for 250,000 iterations, we obtained a \emph{mean absolute error
(MAE)} of 2.95 px on our test set. 
After additional inspection of rendered
videos with label predictions, visual inspection of tracking performance showed
it to be satisfactory. 

The DNN provides for each detected label a likelihood.
Following further inspection of all likelihood plots, we settled on \(p_{cutoff} = 0.9\)
for a successful detection. In addition, detection performance was validated by
examining pixel differences of predicted labels in consecutive frames. Given
the continuity of video footage, these ended up, as expected, close to zero.
Out-of-sample generalization was not necessary,
due to all selected training frames stemming from the same distribution as the
test set data.

%% file: tables/cascade.tex
\begin{tabular}{lr}
\toprule
{} &  p(gaze cascade) \\
participant &                  \\
\midrule
02w20       &           0.9758 \\
04m20       &           0.1249 \\
09w23       &           0.9210 \\
11m26       &           0.3636 \\
13w23       &           0.3476 \\
20m31       &           0.4784 \\
21w20       &           0.6216 \\
26m21       &           0.9673 \\
29w25       &           0.0878 \\
\bottomrule
\end{tabular}

%% file: main.bbl
\begin{thebibliography}{30}
\providecommand{\natexlab}[1]{#1}
\providecommand{\url}[1]{\texttt{#1}}
\providecommand{\urlprefix}{}

\bibitem[{Scott and Dodgson(2022)Scott, Joshua J. and Dodgson, Neil A.}]{scottEvaluatingRealismExampleBased2022}
Scott JJ, Dodgson NA.
\newblock Evaluating {{Realism}} in {{Example-Based Terrain Synthesis}}.
\newblock ACM Transactions on Applied Perception 2022;\urlprefix\url{https://doi.org/10.1145/3531526}.

\bibitem[{Hinde et~al.(2022)Hinde, Stephen J. and Noland, Katy C. and Thomas, Graham A. and Bull, David R. and Gilchrist, Iain D.}]{hindeImmersivePropertiesHigh2022}
Hinde SJ, Noland KC, Thomas GA, Bull DR, Gilchrist ID.
\newblock On the {{Immersive Properties}} of {{High Dynamic Range Video}}.
\newblock ACM Transactions on Applied Perception 2022;\urlprefix\url{https://doi.org/10.1145/3524692}.

\bibitem[{Almutairi et~al.(2021)Almutairi, Aeshah and Ivrissimtzis, Ioannis and Saarela, Toni}]{almutairiImperceptibilityThresholdsQuantised2021}
Almutairi A, Ivrissimtzis I, Saarela T.
\newblock Imperceptibility Thresholds in Quantised {{3D}} Triangle Meshes.
\newblock In: 2021 {{The}} 4th {{International Conference}} on {{Image}} and {{Graphics Processing}} {{ICIGP}} 2021, {Association for Computing Machinery}; 2021. p. 131--136.
\newblock \urlprefix\url{https://doi.org/10.1145/3447587.3447606}.

\bibitem[{Fechner(1860)Fechner, Gustav Theodor}]{fechnerElementePsychophysik1860}
Fechner GT.
\newblock Elemente Der Psychophysik, vol.~2.
\newblock {Breitkopf u. Härtel}; 1860.

\bibitem[{Field(1987)Field, David J.}]{fieldRelationsStatisticsNatural1987}
Field DJ.
\newblock Relations between the Statistics of Natural Images and the Response Properties of Cortical Cells.
\newblock Journal of the Optical Society of America A 1987;4(12):2379.
\newblock \urlprefix\url{https://www.osapublishing.org/abstract.cfm?URI=josaa-4-12-2379}.

\bibitem[{Rideaux and Welchman(2020)Rideaux, Reuben and Welchman, Andrew E.}]{rideauxStillItMoves2020}
Rideaux R, Welchman AE.
\newblock But {{Still It Moves}}: {{Static Image Statistics Underlie How We See Motion}}.
\newblock Journal of Neuroscience 2020;40(12):2538--2552.
\newblock \urlprefix\url{https://www.jneurosci.org/content/40/12/2538}.

\bibitem[{Knopp et~al.(2019)Knopp, Benjamin and Velychko, Dmytro and Dreibrodt, Johannes and Endres, Dominik}]{knoppPredictingPerceivedNaturalness2018}
Knopp B, Velychko D, Dreibrodt J, Endres D.
\newblock Predicting perceived naturalness of human animations based on generative movement primitive models.
\newblock ACM Transactions on Applied Perception (TAP) 2019;16(3):1--18.

\bibitem[{McGuigan(2006)McGuigan, Michael D.}]{mcguiganGraphicsTuringTest2006}
McGuigan MD.
\newblock Graphics {{Turing Test}}.
\newblock CoRR 2006;abs/cs/0603132.
\newblock \urlprefix\url{http://arxiv.org/abs/cs/0603132}.

\bibitem[{Shimojo et~al.(2003)Shimojo, Shinsuke and Simion, Claudiu and Shimojo, Eiko and Scheier, Christian}]{shimojoGazeBiasBoth2003}
Shimojo S, Simion C, Shimojo E, Scheier C.
\newblock Gaze Bias Both Reflects and Influences Preference.
\newblock Nature Neuroscience 2003;6(12):1317--1322.
\newblock \urlprefix\url{https://www.nature.com/articles/nn1150}.

\bibitem[{Krajbich et~al.(2010)Krajbich, Ian and Armel, Carrie and Rangel, Antonio}]{krajbichVisualFixationsComputation2010}
Krajbich I, Armel C, Rangel A.
\newblock Visual Fixations and the Computation and Comparison of Value in Simple Choice.
\newblock Nature Neuroscience 2010;13(10):1292--1298.
\newblock \urlprefix\url{https://www.nature.com/articles/nn.2635}.

\bibitem[{Taubert et~al.(2012)Taubert, Nick and Christensen, Andrea and Endres, Dominik and Giese, Martin A.}]{taubertOnlineSimulationEmotional2012}
Taubert N, Christensen A, Endres D, Giese MA.
\newblock Online Simulation of Emotional Interactive Behaviors with Hierarchical {{Gaussian}} Process Dynamical Models.
\newblock In: Proceedings of the {{ACM Symposium}} on {{Applied Perception}} - {{SAP}} '12 {ACM Press}; 2012. p.~25.
\newblock \urlprefix\url{http://dl.acm.org/citation.cfm?doid=2338676.2338682}.

\bibitem[{Chiovetto et~al.(2018)Chiovetto, Enrico and Curio, Cristóbal and Endres, Dominik and Giese, Martin A.}]{chiovettoPerceptualIntegrationKinematic2018}
Chiovetto E, Curio C, Endres D, Giese MA.
\newblock Perceptual Integration of Kinematic Components in the Recognition of Emotional Facial Expressions.
\newblock Journal of Vision 2018;18(4):13--13.
\newblock \urlprefix\url{https://jov.arvojournals.org/article.aspx?articleid=2678770}.

\bibitem[{Calbi et~al.(2021)Calbi, M. and Langiulli, N. and Siri, F. and Umiltà, M. A. and Gallese, V.}]{calbiVisualExplorationEmotional2021}
Calbi M, Langiulli N, Siri F, Umiltà MA, Gallese V.
\newblock Visual Exploration of Emotional Body Language: A Behavioural and Eye-Tracking Study.
\newblock Psychological Research 2021;85(6):2326--2339.
\newblock \urlprefix\url{https://doi.org/10.1007/s00426-020-01416-y}.

\bibitem[{Geangu and Vuong(2020)Geangu, Elena and Vuong, Quoc}]{geanguLookBodyEyetracking2020}
Geangu E, Vuong Q.
\newblock Look up to the Body : An Eye-Tracking Investigation of 7-Months-Old Infants' Visual Exploration of Emotion Body Expressions.
\newblock Infant Behavior and Development 2020;\urlprefix\url{https://doi.org/10.1016/j.infbeh.2020.101473}.

\bibitem[{Morita et~al.(2012)Morita, Tomoyo and Slaughter, Virginia and Katayama, Nobuko and Kitazaki, Michiteru and Kakigi, Ryusuke and Itakura, Shoji}]{moritaInfantAdultPerceptions2012a}
Morita T, Slaughter V, Katayama N, Kitazaki M, Kakigi R, Itakura S.
\newblock Infant and Adult Perceptions of Possible and Impossible Body Movements: {{An}} Eye-Tracking Study.
\newblock Journal of Experimental Child Psychology 2012;113(3):401--414.
\newblock \urlprefix\url{https://linkinghub.elsevier.com/retrieve/pii/S0022096512001282}.

\bibitem[{Matarić and Pomplun(1998)Matarić, Maja J and Pomplun, Marc}]{mataricFixationBehaviorObservation1998a}
Matarić MJ, Pomplun M.
\newblock Fixation Behavior in Observation and Imitation of Human Movement.
\newblock Cognitive Brain Research 1998;7(2):191--202.
\newblock \urlprefix\url{https://linkinghub.elsevier.com/retrieve/pii/S0926641098000251}.

\bibitem[{Pettersson and Falkman(2020)Pettersson, Julius and Falkman, Petter}]{petterssonHumanMovementDirection2020}
Pettersson J, Falkman P.
\newblock Human {{Movement Direction Classification}} Using {{Virtual Reality}} and {{Eye Tracking}}.
\newblock Procedia Manufacturing 2020;51:95--102.
\newblock \urlprefix\url{https://www.sciencedirect.com/science/article/pii/S2351978920318709}.

\bibitem[{Pettersson and Falkman(2021)Pettersson, Julius and Falkman, Petter}]{petterssonHumanMovementDirection2021}
Pettersson J, Falkman P.
\newblock Human {{Movement Direction Prediction}} Using {{Virtual Reality}} and {{Eye Tracking}}.
\newblock In: 2021 22nd {{IEEE International Conference}} on {{Industrial Technology}} ({{ICIT}}), vol.~1; 2021. p. 889--894.

\bibitem[{Johansson(1973)Johansson, Gunnar}]{johanssonVisualPerceptionBiological1973}
Johansson G.
\newblock Visual Perception of Biological Motion and a Model for Its Analysis.
\newblock Perception \& Psychophysics 1973;14(2):201--211.
\newblock \urlprefix\url{http://www.springerlink.com/index/10.3758/BF03212378}.

\bibitem[{Mathis et~al.(2018)Alexander Mathis and Pranav Mamidanna and Kevin M. Cury and Taiga Abe and Venkatesh N. Murthy and Mackenzie W. Mathis and Matthias Bethge}]{Mathisetal2018}
Mathis A, Mamidanna P, Cury KM, Abe T, Murthy VN, Mathis MW, et~al.
\newblock DeepLabCut: markerless pose estimation of user-defined body parts with deep learning.
\newblock Nature Neuroscience 2018;\urlprefix\url{https://www.nature.com/articles/s41593-018-0209-y}.

\bibitem[{Nath et~al.(2019)Nath, Tanmay and Mathis, Alexander and Chen, An Chi and Patel, Amir and Bethge, Matthias and Mathis, Mackenzie W}]{NathMathisetal2019}
Nath T, Mathis A, Chen AC, Patel A, Bethge M, Mathis MW.
\newblock Using DeepLabCut for 3D markerless pose estimation across species and behaviors.
\newblock Nature Protocols 2019;\urlprefix\url{https://doi.org/10.1038/s41596-019-0176-0}.

\bibitem[{He et~al.(2015)Kaiming He and Xiangyu Zhang and Shaoqing Ren and Jian Sun}]{he2015deep}
He K, Zhang X, Ren S, Sun J, Deep Residual Learning for Image Recognition; 2015.

\bibitem[{Insafutdinov et~al.(2016)Eldar Insafutdinov and Leonid Pishchulin and Bjoern Andres and Mykhaylo Andriluka and Bernt Schiele}]{insafutdinov2016deepercut}
Insafutdinov E, Pishchulin L, Andres B, Andriluka M, Schiele B, DeeperCut: A Deeper, Stronger, and Faster Multi-Person Pose Estimation Model; 2016.

\bibitem[{Andriluka et~al.(2014)Mykhaylo Andriluka and Leonid Pishchulin and Peter Gehler and Schiele, Bernt}]{andriluka14cvpr}
Andriluka M, Pishchulin L, Gehler P, Schiele B.
\newblock 2D Human Pose Estimation: New Benchmark and State of the Art Analysis.
\newblock In: IEEE Conference on Computer Vision and Pattern Recognition (CVPR); 2014. .

\bibitem[{Rayner(2009)Rayner, Keith}]{raynerEyeMovementsReading}
Rayner K.
\newblock Eye movements in reading: Models and data.
\newblock Journal of eye movement research 2009;2(5):1.

\bibitem[{Salvatier et~al.(2016)Salvatier, John and Wiecki, Thomas V. and Fonnesbeck, Christopher}]{salvatierProbabilisticProgrammingPython2016}
Salvatier J, Wiecki TV, Fonnesbeck C.
\newblock Probabilistic Programming in {{Python}} Using {{PyMC3}}.
\newblock PeerJ Computer Science 2016;2:e55.
\newblock \urlprefix\url{https://peerj.com/articles/cs-55}.

\bibitem[{Cover(1999)Cover, Thomas M.}]{coverElementsInformationTheory1999}
Cover TM.
\newblock Elements of Information Theory.
\newblock {John Wiley \& Sons}; 1999.

\bibitem[{Gronau et~al.(2017)Gronau, Quentin F. and Sarafoglou, Alexandra and Matzke, Dora and Ly, Alexander and Boehm, Udo and Marsman, Maarten and Leslie, David S. and Forster, Jonathan J. and Wagenmakers, Eric-Jan and Steingroever, Helen}]{gronauTutorialBridgeSampling2017}
Gronau QF, Sarafoglou A, Matzke D, Ly A, Boehm U, Marsman M, et~al.
\newblock A Tutorial on Bridge Sampling.
\newblock Journal of Mathematical Psychology 2017;81:80--97.
\newblock \urlprefix\url{https://www.sciencedirect.com/science/article/pii/S0022249617300640}.

\bibitem[{Araujo et~al.(2001)Araujo, Christian and Kowler, Eileen and Pavel, Misha}]{araujoEyeMovementsVisual2001}
Araujo C, Kowler E, Pavel M.
\newblock Eye Movements during Visual Search: {{The}} Costs of Choosing the Optimal Path.
\newblock Vision research 2001;41(25-26):3613--3625.

\bibitem[{Agtzidis(2020)Agtzidis, Ioannis}]{agtzidisBetterUnderstandingEye}
Agtzidis I, Towards a Better Understanding of Eye Movements in Natural Contexts; 2020.

\end{thebibliography}
